       \newcommand{\nl}{\nonumber \\}
       \newcommand{\beq}{\begin{equation}}
       \newcommand{\eeq}{\end{equation}}
       \newcommand{\beqa}{\begin{eqnarray}}
       \newcommand{\eeqa}{\end{eqnarray}}
       \newcommand{\beqas}{\begin{eqnarray*}}
       \newcommand{\eeqas}{\end{eqnarray*}}
       
       \newcommand{\ba}{{\mathbf a}}
       \newcommand{\bb}{{\mathbf b}}

       \newcommand{\bj}{{\mathbf j}}

       \newcommand{\br}{{\mathbf r}}

       \newcommand{\bv}{{\mathbf v}}

       \newcommand{\bA}{{\mathbf A}}

       \newcommand{\bF}{{\mathbf F}}
       \newcommand{\bG}{{\mathbf G}}

       \newcommand{\bxi}{\mbox{\boldmath $\xi$}}
       \newcommand{\dadb}[2]{\frac{{  d}#1}{{  d}#2}}

       \newcommand{\parparb}[1]{\frac{\partial}{\partial #1}}
       \newcommand{\paraparb}[2]{\frac{\partial #1}{\partial #2}}

      \newcommand{\dbxi}{\dot{\bxi}}  
      \newcommand{\dbA}{\dot{\bA}}        

\documentclass[twocolumn,amsmath,amssymb,aps]{revtex4-1}

\usepackage[utf8]{inputenc}
\usepackage[T1]{fontenc}
\usepackage{amsmath}

\usepackage{mathtools}
\usepackage{amsfonts}
\usepackage{graphicx}
\usepackage{dcolumn}
\usepackage{bm}
\usepackage[export]{adjustbox}
\usepackage{wrapfig}
\usepackage{lipsum}
\usepackage[titletoc,toc,title]{appendix}
\usepackage[colorlinks]{hyperref}
\usepackage{color}

\begin{document}

\title{Comment on ``Pfirsch-Tasso versus standard approaches in the plasma stability theory
including the resistive wall effects" [Phys. Plasmas 24, 112513 (2017)]}

\author{H. Tasso$^1$}
 \email{hetasso@t-online.de}
\author{G. N. Throumoulopoulos$^2$}%
 \email{gthroum@uoi.gr}
\affiliation{
$^1$Max-Planck-Institut f$\ddot{\mbox{u}}$r Plasmaphysik, Boltzmannstrasse 2, 85748 Garching, Germany  \\
$^2$Department of Physics, University of Ioannina, GR 451 10 Ioannina, Greece  
}%

\date{\today}

\begin{abstract}
 In the commended paper it is claimed that the proves of the ``Resitive-Wall-Mode theorem'' by Pfirsch and Tasso [Nucl. Fusion \textbf{11}, 259 (1971)] and extensions of that theorem for time dependent wall resistivity and equilibrium plasma flow are not detailed and that there are limitations restricting their  applicability. In response, we provide here pertinent detailed derivations showing that the proves of the above mentioned theorems are rigorous and complete, unlike the considerations in the commended paper which ignore the self adjointness of the operator $\nabla\times\nabla\times$ and the fact that the force  operator in the linearized ideal MHD momentum equation remains self adjoint in the presence of equilibrium flows. As a matter of fact it is proved here that, because of the self adjointness of the operator $\nabla\times\nabla\times$, a claimed in the commended paper additional term in Ohm's law  vanishes identically. 
\end{abstract}

\pacs{Valid PACS appear here}
\maketitle



The subject of \cite{pfta} is the following theorem: ``An MHD-unstable configuration with a dissipationless plasma surrounded by vacuum and possibly superconducting walls can not be stabilized by introducing walls of finite electrical conductivity''. The pertinent instability is known nowadays as resistive wall  mode (RWM). Later the prove of the theorem was extended for walls with time dependent electrical conductivity \cite{tath1} and for stationary ideal MHD equilibria with equilibrium flows \cite{tath2}.  In \cite{pu} it is stated that the main relations in \cite{pfta} have been given there without detailed proves and that there are unknown limitations that restrict the applicability of the Pfisrch-Tasso variational principle,  which
is claimed to be restored  in the framework of a generic plasma model. In view of that criticism we first will give  here  in detail the proves in \cite{pfta}-\cite{tath2}. 

 The  theorem is based on the linearized momentum equation for the  displacement vector $\bxi$ in an ideal MHD plasma about a static equilibrium given  by
 \begin{equation}
\rho_{0}  \ddot{\mbox{\boldmath $\xi$}} - {\bf F}[\mbox{\boldmath
$\xi$}]=0
\label{me1}
\end{equation}
in connection with the well known energy principle \cite{befr}.  Here, 
$\rho_0$ is the equilibrium mass density and ${\bf F}$ is the standard  self-adjoint force operator (which corresponds to $-\bF$ in \cite{pfta}).
In the region outside the plasma the perturbation for the vector potential, $\bA$, after fixing the gauge  by annihilating  the scalar potential,   is
governed by Ohm's law
\begin{equation}
\label{Ohm}
\nabla\times\nabla\times {\bf A} = - \sigma  \dot {\bf A}
\end{equation}
where $\sigma$ is the conductivity in the outer region. 
It is zero for vacuum regions and has some finite value for the inserted walls. The walls are assumed not to  touch the plasma. 
At the interface between the plasma and the outer region, the following boundary condition should be satisfied:
\begin{equation}
{\bf n}\times {\bf A} = - ({\bf n}\cdot {\mbox{\boldmath $\xi$}})
{\bf B_{0}}
\label{bc}
\end{equation}
where ${\bf B_{0}}$ is the equilibrium magnetic field at the
vacuum side of the interface whose normal vector is ${\bf n}$.
(\ref{bc}) is derived on page 278 of \cite{go1}.
The scalar product of (\ref{me1}) with $\dot{\bxi}$ on account of self-adjointness of $\bF$ yields
\beqa
 \left(\rho_0 \dbxi, \ddot{\mbox{\boldmath $\xi$}}\right)_p - (\dbxi, \bF[\bxi])_p  & & \nl
= \frac{1}{2}\dadb{\left(\rho_0 \dbxi,\dbxi\right)_p}{t}
-\frac{1}{2}\left(\dbxi, \bF[\bxi]\right)_p
-\frac{1}{2}\left(\bxi, \bF[\dbxi]\right)_p\nl 
=\frac{1}{2}\dadb{(\rho_0 \dbxi,\dbxi)_p}{t} -\frac{1}{2}\dadb{\left(\bxi, \bF[\xi]\right)_p}{t}= 0, 
\label{spr1}
\eeqa
where
the scalar product is defined as
$$
\left(\ba, \bb\right)_p:=\int_{p}\, \ba\cdot\bb\,  d^3 x
$$
with the subscript $p$  indicating the plasma region. In a similar way, the scalar product of  (\ref{Ohm}) with $\dbA$,  on account of
the  self-adjointness of the operator $\nabla\times\nabla \times $ and  the condition that $\bA$ has to vanish at infinity or be normal to potential superconducting walls,   yields
\beq
\label{spr2}
\frac{1}{2} \dadb{\left(\bA, \nabla\times\nabla\times \bA \right)_{or}}{t}=-\sigma\left(\dbA,\dbA\right)_{or} 
\eeq
with the subscript $or$ indicating the outer plasma region.
Adding the last of Eqs. (\ref{spr1}) to  (\ref{spr2}) we get the relation
\beq
\label{sst}
\dadb{(\left(K+W\right)}{t}= -\sigma\left(\dbA,\dbA\right)_{or}
\eeq
where
\beq
\label{energies}
K:=\frac{1}{2}\dadb{\left(\rho_0 \dbxi,\dbxi\right)_p}{t},\ \  W:=\frac{-\left(\bxi, \bF[\bxi]\right)_p+ \left(\bA, \nabla\times\nabla\times \bA \right)_{or}}{2}
\eeq
The last relation  implies that the  quantity $K+W$  has a negative time derivative. This allows
us to apply the second method of Lyapunov, which states that if $K + W$ has no
definite sign, then the system is "Lyapunov unstable". Note that $K + W$ is
indefinite in sign if the system is MHD unstable in the absence of resistive walls. Although in the above prove the boundary condition (\ref{bc}) was  not employed explicitly, it is needed to prove the self-adjointness of $\bF$. This completes the prove of the RWM-theorem of \cite{pfta}. In \cite{tath1} we just observed that (\ref{sst}) remains valid for time dependent conductivities, $\sigma(t)$, thus extending the  theorem to walls with time dependent resistivities. 

Coming now to stationary equilibria, i.e. ideal MHD steady states  with flow examined in \cite{tath2}, the linearized momentum equation for the displacement vector $\bxi$ becomes \cite{frro}
\begin{equation}
\label{me2}
\rho_{0}  \ddot{\mbox{\boldmath $\xi$}} + 2\rho_{0}{\bf
v_{0}}\cdot\nabla\dot{\mbox{\boldmath $\xi$}} - {\bf
G}[\mbox{\boldmath$\xi$}] = 0.
\end{equation}
Here,  $\bv_0$ is the equilibrium fluid-element velocity and $\bG$ the generalized force operator: 
\beq
\label{gfo}
\bG[\bxi]=\bF[\bxi]+ \nabla\cdot \left(\bxi\rho_0\bv\cdot\nabla\bv-\rho_0\bv\bv\cdot\nabla\bxi\right),
\eeq
involving the standard force operator, $\bF$,  of static equilibria.
Derivations of (\ref{me2}) and (\ref{gfo}), which are identical with Eqs.    (1) .\cite{tath2} and (2).\cite{tath2} with $\bF$ therein  corresponding to $-\bG$ here, are provided in Subsection 12.2.2 of \cite{go2}.   As  pointed out in \cite{go2}, because of the flow terms in (\ref{gfo})   there has been a misunderstanding in the literature,  including \cite{pu},  that the operator $\bG$ is non self-adjoint. This is incorrect.  The self adjointnesess of $\bG$ is proved in \cite{go2}. 
Thus, the only term which renders the stability problem tougher in the presence of flow is the term $ 2\rho_{0}{\bf v_{0}}\cdot\nabla\dot{\mbox{\boldmath $\xi$}}$  in (\ref{me2}) coming from the convective derivative. Because of the  skew-symmetry of $\rho_{0}{\bf v_{0}}\cdot\nabla$, called gyroscopic operator, one would expect that the skew symmetric  term in (\ref{me2}) does not contribute to the variational principle, i.e. that it holds $(\dbxi, \rho_0\bv_0\cdot\nabla \dbxi)_p=0$. The proof of this conjecture is provided in \cite{tath2}; therefore,  in the presence of equilibrium flow Eq. (\ref{sst}) keeps valid (with $\bF$ therein replaced by $\bG$) and the system remains Lyapunov unstable. 

Unlike  what is stated in \cite{pu} the proofs in \cite{pfta}-\cite{tath2} hold for arbitrary perturbations $\bxi$ and not only for exponential in time ones, i.e. perturbations of the form $\bxi(\br,t)=\tilde{\bxi}(\br)\exp(i\gamma t)$. However, the  Lyapunov method does not provide any information on the growth rate of the instability which may be of  practical importance.  This was the motivation in \cite{pfta} to further consider exponential in time perturbations in order to make an estimate on the growth rate.
 It may be noted that such perturbations are the standard ones adopted in the literature, since they are related with the fastest growth rates;  they  reduce  the linear stability problems  to  initial value ones, which are the basis of  spectral theory (cf. Chapter 6 in \cite{go1} for static equilibria and Chapter 12 of \cite{go2} for stationary ones).  This additional consideration led to   relation (18).\cite{pfta},  regarding  the upper bound for $\gamma$. Such growth-rate estimations  have not been made in \cite{tath1} and \cite{tath2}, a fact which is clarified even in the Abstract of \cite{tath1}, verbatim: ``A previous theorem by Pfisrch and Tasso \cite{pfta} concerning ``resistive wall modes'' is extended in a `weak sense' to the case of time-depended wall resistivity'',  that is in the sense that for time-dependent wall resistivity the modes remains Lyapunov unstable.

In \cite{pu} the author pursues to retain the form of the linearized momentum equation  (\ref{me1}) for any plasma model, which is claimed that  
can include  dissipative or kinetic effects,  by just replacing the operator $\bF$ by a ``generic  operator'', $\bF_P$ (denoted by the  symbol $\bF$ therein),  describing those additional effects.  $\bF_p$ is never specified but it is employed just  as a symbol. For static ideal MHD equilibria $\bF_P$ is equal to $\bF$ but for stationary ones it holds $\bF_P=\bG-2\rho_0\bv_0\cdot\nabla\dbxi$ by Eq. (\ref{me2}); thus for stationary ideal MHD equlibria,  $\bF_p$ consists of a superposition of the self adjoint operator $\bG$ and the skew-symmetric one,  $-2\rho_0\bv_0\nabla\cdot\dbxi$.  The introduction of $\bF_P$ hides this information; thus,  in Section VI of \cite{pu}, $\bF_P$ is decomposed as $\bF_P=\bF_{id}+\bF_{non}$ with $\bF_{id}$ representing the standard self-adjoint force operator for static equilibria and $\bF_{non}$, never specified, representing flow and non ideal effects of the model. Since now $\bF_P$ is not self adjoint,  it gives rise to a second term, $I_{p}$,  on the rhs of  (\ref{sst}) given by
\beq
\label{pu1}
I_{p}:=\frac{1}{2}\int_{p}\, \left(\dbxi\cdot\bF_P[\bxi]-\bxi\cdot \bF_P[\dbxi]\right)\, d^3 x
\eeq
(cf. Eqs. (14).\cite{pu} and (20).\cite{pu}). However, as we have proved above  in the framework of ideal MHD,  $I_{p}$ vanishes for static as well stationary equilibria examined in \cite{pfta}-\cite{tath2}. More generally, $I_{pl}$ vanishes if the operator $\bF_{non}$ is skew symmetric, then  implying  that the system remains Lyapunov unstable. 

Furthermore,  in \cite{pu} the self adjointness of the operator
 $\nabla\times\nabla\times$  
in (\ref{Ohm})
 is ignored; specifically,  the author introduces a perturbed current density, $\tilde{\bj}=\nabla\times\nabla\times \bA$ 
(cf. Eq. (7).\cite{pu}), thus hiding  the self adjointness of  that operator. Subsequently,  instead of taking the scalar product of equation $\rho_0\partial^2\bxi/\partial t^2= \bF_P[\bxi]$ 
with $\dbxi$,  the author  takes the  scalar product of that equation with $\bxi$ and calculates the time derivative of the quantity $K +W$ explicitly by using the definitions (\ref{energies}) for $K$ and $W$. This procedure makes  the further analysis in \cite{pu}  complicated and gives rise to another  term $I_{or}$ on the rhs of (\ref{sst}) (cf. Eq. (20).\cite{pu})
\beq
I_{or}:=\frac{1}{2}\int_{or}\, \left(\bA\cdot\paraparb{\tilde{\bj}}{t}-\paraparb{\bA}{t}\cdot\tilde{\bj}\right)\, d^3 x
\eeq
(first of Eqs. (21).\cite{pu}). However, because of the definition $\tilde{\bj}=\nabla\times\nabla \times \bA$  and the self-adjointness of the operator  $\nabla\times\nabla\times$ this term vanishes identically irrespective of the plasma model:
\beqa
I_{or}&=& \frac{1}{2}\int_{or}\, \left(\bA\cdot\paraparb{\tilde{\bj}}{t}-\paraparb{\bA}{t}\cdot\tilde{\bj}\right)\, d^3 x \nl 
& &=\frac{1}{2}\int_{or}\,\left(\bA\cdot\parparb{t}\nabla\times\nabla\times\bA - \dot{\bA}\cdot\tilde{\bj}\right)\, d^3  x \nl
& & =  \frac{1}{2}\int_{or}\, \left(\bA\cdot \nabla\times\nabla\times \dot{\bA}-\dot{\bA}\cdot\tilde{\bj}\right)\, d^3 x \nl 
& &= \frac{1}{2}\int_{or}\, \left(\nabla\times\nabla\times\bA \cdot  \dot{\bA}-\dot{\bA}\cdot\tilde{\bj}\right)\, d^3 x \nl
& &=  \frac{1}{2}\int_{or}\, \left(\tilde{\bj}\cdot\dot{\bA}- \dot{\bA}\cdot\tilde{\bj}\right)\, d^3 x \equiv 0
 \eeqa
 Therefore, Eq. (\ref{spr2})  remains valid  for any plasma model associated with $\bF_P$,  when  $\bF_P$ can be written as a superposition of a symmetric operator and a skew-symmetric one. 

Finally, we would like to note that in order that a stability  study leads to reliable conclusions of practical importance, e.g. in connection with the ITER project as claimed in \cite{pu}, it should fulfill a couple of requirements.  First, it should be performed within the framework of a concrete,  well defined plasma model and second the force operator in the momentum equation should be derived concretely and self consistently. The latter requirement,   which needs the self consistent use of  the whole set of equations of the model,   remains in general an open tough question, in particular if the model involves dissipative or kinetic effects. Both of the above requirements are  well fulfilled in \cite{pfta}-\cite{tath2}, which are based on the  pioneering  ideal MHD papers \cite{befr} and \cite{frro}, but they are not fulfilled
in \cite{pu}. 

 This  study was performed within the framework of the EUROfusion Consortium and
has received funding from the National Program for the Controlled Thermonuclear
Fusion, Hellenic Republic. The views and opinions expressed herein do not necessarily
reflect those of the European Commission.

\end{document}